\documentclass[aps,prl,twocolumn,superscriptaddress]{revtex4}
\usepackage{graphicx,amsmath}
\usepackage[amssymb]{SIunits}
\usepackage[usenames,dvipsnames]{color}   
\usepackage{color}
\definecolor{orange}{rgb}{1,0.6,0}

\begin{document}

\title{Logarithmic boundary layers in highly turbulent Taylor-Couette flow}

\author{Sander G. Huisman}  
\affiliation{Department of Applied Physics and J. M. Burgers Centre for Fluid Dynamics, University of Twente, P.O. Box 217, 7500 AE Enschede, The Netherlands}
\author{Sven Scharnowski}
\author{Christian Cierpka}
\author{Christian J. K\"ahler}
\affiliation{Institut f\"ur Str\"omungsmechanik und Aerodynamik, Universit\"at der Bundeswehr M\"unchen, Werner-Heisenberg-Weg 39, 85577 Neubiberg, Germany}
\author{Detlef Lohse}
\author{Chao Sun}

\affiliation{Department of Applied Physics and J. M. Burgers Centre for Fluid Dynamics, University of Twente, P.O. Box 217, 7500 AE Enschede, The Netherlands}

\date{\today}

\begin{abstract} 
We provide direct measurements of the boundary layer properties in highly turbulent Taylor-Couette flow up to $\text{Ta}=6.2 \times 10^{12}$ using high-resolution particle image velocimetry (PIV). We find that the mean azimuthal velocity profile at the inner and outer cylinder can be fitted by the von K\'arm\'an log law $u^+ = \frac 1\kappa \ln y^+ +B$. The von K\'arm\'an constant $\kappa$ is found to depend on the driving strength $\text{Ta}$ and for large $\text{Ta}$ asymptotically approaches $\kappa \approx 0.40$. The variance profiles of the local azimuthal velocity have a universal peak around $y^+ \approx 12$ and collapse when rescaled with the driving velocity (and not with the friction velocity), displaying a log-dependence of $y^+$ as also found for channel and pipe flows \cite{marusic2013,meneveau2013}.
\end{abstract}

\maketitle

\begin{figure}[h!t]
	\begin{center}
		\includegraphics{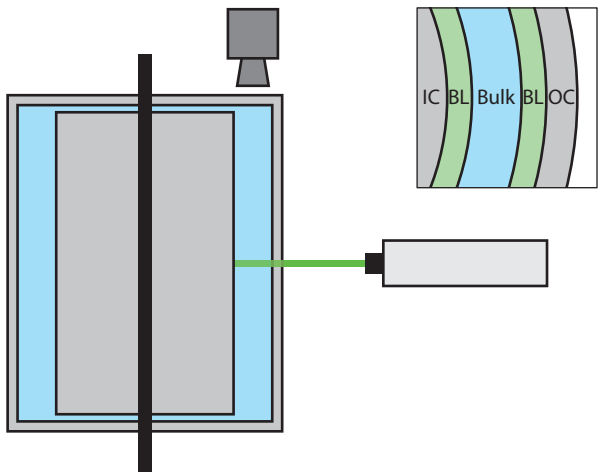}
		\caption{(color online) Sketch of the vertical cross section of the $\text{T}^3\text{C}$ \cite{gil11a}. The flow is illuminated from the side in the horizontal plane using a laser and the flow is imaged from the top using a high-resolution camera. Top-right inset: Schematic top view of different regions inside the gap: \textsf{IC}  (inner cylinder), \textsf{OC} (outer cylinder), and \textsf{BL} (boundary layer). Measurements are done at middle height.}
		\label{fig:setup}
	\end{center} 
\end{figure}

Taylor-Couette (TC) flow is one of the paradigmatical flows in physics of fluids, next to Rayleigh-B\'enard convection, channel flow, and pipe flow. These flows have always been used in order to experimentally test new concepts in fluid dynamics. The TC system consists of two rotating coaxial cylinders shearing a fluid in between the cylinders, see fig.\ \ref{fig:setup}. The Taylor number $\text{Ta}$ quantifies the driving of this system and is directly related to the shear. For increasing $\text{Ta}$ the system is first dominated by coherent structures \cite{and86} whose length scale is of similar size as the gap width. For further increasing $\text{Ta}$ turbulence develops in the bulk at length scales between the integral and the Kolmogorov scale while the boundary layers (BL) are still laminar. This regime, in which the flow has a turbulent bulk and the boundary layers are of Prandtl-Blasius type, is called the classical regime \cite{gro00}, and is analog to the situation in RB convection. By further increasing the Taylor number the system enters the so-called ultimate turbulent state in which also the boundary layers have turned turbulent \cite{kra62,gro11,gro12}. These turbulent states are found to be remarkably similar for RB convection \cite{eck07b}. Based on global transport measurements, the ultimate regime of turbulence sets in at Ra $\sim 10^{14}$ for RB \cite{he12}, but already at Ta $\sim 10^8$ for TC \cite{lat92,lew99,gil12}, due to the fact that mechanical driving in TC is much more efficient in triggering BL instabilities than thermal driving is in RB. 

The coexistence of a laminar type boundary layer and turbulent bulk in classical turbulent RB convection has been well established from numerous experimental and numerical investigations \cite{sun06,sun08,ahl09,loh10,zho10} and theoretical analysis \cite{gro00}. This is also the case for the classical regime in TC flow \cite{wen33,smi82,wereley1998,lew99,don07} and the transition regime to ultimate TC flow \cite{hou11,ost12,brauckmann2013}. Very recently, a direct measurement \cite{ahlers2012} of the mean temperature profile close to the wall in the ultimate RB state revealed logarithmic behavior in the ultimate regime. Neither for TC nor for RB there had been any direct and systematical measurement of the velocity boundary layer in the highly turbulent ultimate state due to the experimental difficulties. However, the boundary layer properties are crucial to understand and define the general picture of the ultimate turbulent state. 
Assuming a logarithmic velocity profile in the boundary layers for highly turbulent TC flow, and matching the mean velocities at midgap, Lathrop {\it et al.} \cite{lat92,lat92pra} obtained a dependence of the global torque and the Reynolds number, which agrees well with the torque measurements in ultimate turbulence regime \cite{lat92,lat92pra,hui12}. Only recently, direct measurements on BLs were conducted by van Hout \& Katz \cite{hou11} for Taylor number up to $2 \times 10^9$, where they focused on the effect of counter-rotation and found that the von K\'arm\'an constant depends on the angular velocity ratio.

In this letter we report the direct systematical experimental investigation of the boundary layer properties for very high Taylor numbers from $\text{Ta} = 9.9 \times 10^8$ to $\text{Ta} = 6.2 \times 10^{12}$ using high resolution PTV and PIV \cite{kah12a,kah12b,sch12} with an unprecedented spatial resolution down to \unit{\approx10}{\micro \meter}. We focus on the case of inner cylinder rotation, and examine the boundary layer properties as a function of $\text{Ta}$ in the ultimate turbulent TC regime.

In analogy to RB convection, in ref. \cite{eck07b} it was mathematically found from the Navier-Stokes and continuity equations that the following quantity is strictly conserved in TC flow:
\begin{align}
	J_\omega &= r^3 \left ( \left \langle u_r \omega \right \rangle_{z,\theta,t} -\nu \partial_r \left \langle \omega \right \rangle_{z,\theta,t} \right) \label{jomega}.
\end{align}
Here $\langle X \rangle_{z,\theta,t}$ represents axial, azimuthal, and time averaging of $X$, $u_r$ is the radial velocity, $\omega$ the angular velocity component $\omega = u_\theta/r$, and $\nu$ is the kinematic viscosity. This flux is made dimensionless by dividing it by the flux for laminar flow: $J_{\omega,\text{lam}}=2\nu r_i^2 r_o^2 (\omega_i - \omega_o)/(r_o^2 - r_i^2)$, giving an angular velocity Nusselt number: $\text{Nu}_\omega = J_\omega / J_{\omega,\text{lam}}$. This transport quantity is independent of $r$; any flux going through an imaginary cylinder with radius $r$ also goes through any other imaginary cylinder, or mathematically $\partial_r \text{Nu}_\omega = 0$. This flux can be measured locally \cite{ji2013,hui12} but also globally \cite{lat92,pao11,gil11,merbold2013} by measuring the torque needed to sustain constant velocity of the cylinders. The torque $\mathcal{T}$ is related to the dimensionless torque $G$ and to $\text{Nu}_\omega$ as follows:
\begin{align}
 G &= \frac{\mathcal{T}}{2\pi \ell \rho \nu^2} = \frac{\text{Nu}_\omega J_{\omega, \text{lam}}}{\nu^2} \label{torquenusselt}
\end{align}
where $\rho$ is the density of the fluid, and $\ell$ the height of the cylinders. We can further relate these quantities to the wall shear stress $\tau_\omega$, the friction velocity $u_\tau$, and the viscous length scale $\delta_\nu$ at the inner cylinder wall:
\begin{align}
 \tau_{\omega,i} &= \frac{\mathcal{T}}{r_i (2 \pi r_i \ell)} \label{eqtaui} \\
 u_{\tau,i} &= \sqrt{\frac{\tau_{\omega,i}}{\rho}} \label{equi} \\
 \delta_{\nu,i} &= \frac{\nu}{u_{\tau,i}} \label{eqdeltai}
\end{align}
Note that as $\text{Nu}_\omega$ is conserved radially, it is the same at both cylinders, and using eq.\ (\ref{torquenusselt}), we see that thus also the torque $\mathcal{T}$ at both cylinders should be the same. Consequently, $\tau_\omega$, $u_\tau$, and $\delta_\nu$ are different at the inner and outer cylinder, and the following relations hold:
\begin{align}
 \tau_{\omega,i} / \tau_{\omega,o} &=  1/ \eta^2 \label{eqtauo} \\
 u_{\tau,i} / u_{\tau,o} &=1 / \eta \label{equo}  \\
 \delta_{\nu,i} / \delta_{\nu,o} &= \eta  \label{eqdeltao}
\end{align}

The apparatus used for the experiments, the Twente turbulent Taylor-Couette ($\text{T}^3\text{C}$), has an inner cylinder with an outer radius of $r_i = \unit{0.200}{\meter}$, a transparent outer cylinder with inner radius $r_o = \unit{0.279}{\meter}$, giving a radius ratio of $\eta=0.716$. The cylinders have a height of $\ell = \unit{0.927}{\meter}$, resulting in an aspect ratio of $\Gamma=\ell/(r_o - r_i) = 11.7$. More details can be found in ref. \cite{gil11a}. In this letter the focus is on inner cylinder rotation only, the outer cylinder remains fixed during all the experiments. For several rotation rates we perform PIV and PTV (particle tracking velocimetry) measurements in our setup. The working fluid (water) is seeded with fluorescent polymer particles \cite{dantec} with a diameter of 1--20\unit{}{\micro \meter}. Using a laser \cite{innolas} we create a horizontal laser sheet of roughly \unit{500}{\micro \meter} thickness that illuminates the particles inside the fluid. The flow is then imaged from the top, see also figure \ref{fig:setup}, using a high-resolution camera \cite{pco} with large dynamical range. In front of the camera a \unit{50}{\milli \meter} objective lens was mounted, with the working distance of approximately \unit{600}{\milli \meter}, giving a scaling factor of \unit{\approx 54}{\micro \meter \per px}. 

\begin{figure}[h!]
	\begin{center}
		\includegraphics{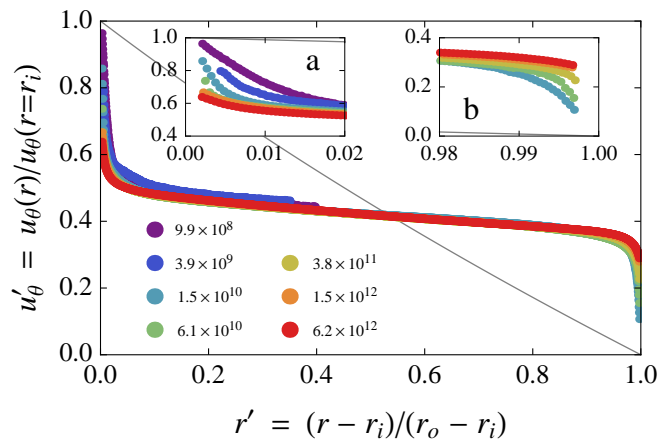}
		\caption{(color online) Azimuthal velocity profiles for varying $Ta$ across the gap of the Taylor-Couette apparatus. The legend indicates the Taylor numbers of the experiments. Insets a and b show a zoom of the data of the inner and outer boundary layer, respectively. Individual data-points are plotted in all 3 figures, showing the high resolution of the measurements. The gray solid line is the exact laminar solution of the Navier-Stokes equations for infinitely long cylinders.}
		\label{fig:fullprofiles}
	\end{center} 
\end{figure}

\begin{figure*}[ht!]
	\begin{minipage}[b]{0.49\textwidth}
		\centering
		\includegraphics{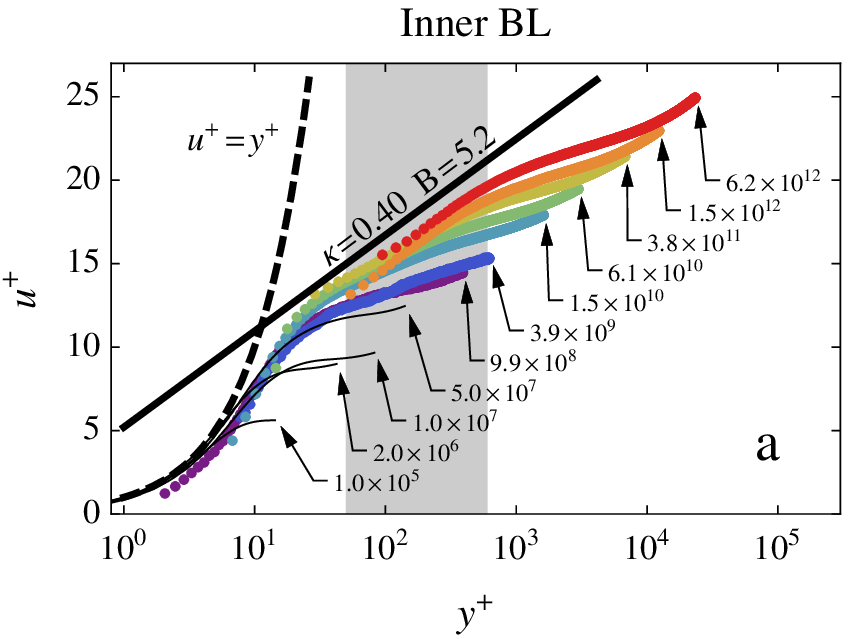}
	\end{minipage}
	\begin{minipage}[b]{0.49\textwidth}
	\centering
		\includegraphics{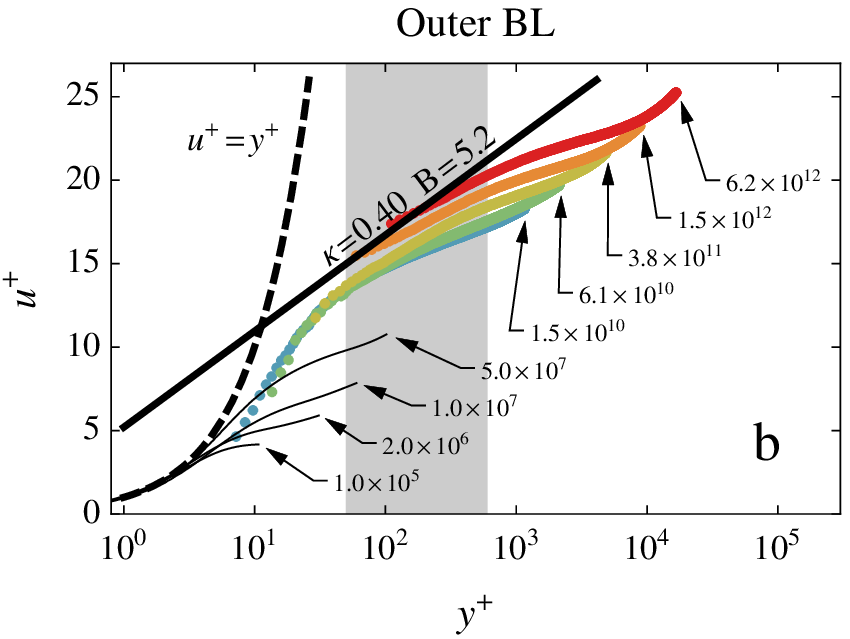}
	\end{minipage}
	\caption{(color online)(a) Azimuthal velocity profile near the inner cylinder ($r' \in [0,1/2]$) for varying $\text{Ta}$. $y^+$ is the distance from the inner cylinder in units of the viscous length scale $\delta_\nu$. $u^+$ is defined as $\big(u(r_i)- u(r)\big)/u_\tau$, where $u(r)$ is the azimuthal component of the velocity and $u(r_i)$ is the azimuthal velocity of the inner cylinder. (b) Azimuthal velocity profile near the outer cylinder $r' \in [1/2,1]$. $y^+=\big(r_o - r\big)/\delta_\nu$ is the distance from the outer cylinder in wall units. For the case of the outer boundary the velocity is scaled as $u^+=u(r)/u_\tau$. Both figures include the logarithmic law of the wall $u^+ = \frac 1\kappa \ln y^+ +B$ by von K\'arm\'an, with the typical values of $\kappa = 0.40$ and $B=5.2$, the viscous boundary layer $u^+ = y^+$, and in gray the fitting domain $y^+ \in [50,600]$. Each data-set is accompanied by an arrow with caption indicating the Taylor number.}
	\label{fig:logprofiles}
\end{figure*}

For each rotational velocity $10^4$ image pairs were acquired at a recording frequency of \unit{10}{\hertz}. The mean velocity distribution was computed using single-pixel ensemble-correlation, leading to a final resolution of \unit{\approx 150}{\micro \meter}, resulting in over 500 independent datapoints in the \unit{80}{\milli \meter} gap \cite{kahler2006,kah12a}. The standard deviation was directly computed from the velocity probability density function, which was extracted from the shape of the correlation function, as discussed in ref. \cite{sch12}. This procedure ensures that all turbulent scales are included in the standard deviation. In contrast to standard PIV analysis, where only spatially low-pass filtered results are achieved, here also the contribution of the small scale fluctuations are properly considered. In order to resolve the near wall-region at the inner cylinder, a microscope \cite{microscope} with a focal length of \unit{300}{\milli \meter} was mounted in front of the camera. With this setup, a scaling factor of \unit{\approx 10}{\micro \meter \per px} was achieved. We evaluate the near-wall region with PTV methods, which is well suited for this purpose \cite{kah12b}.

Figure \ref{fig:fullprofiles} shows all the measured profiles; 5 covering the full gap, and 2 covering just the region near the inner cylinder. We see that the profiles do not conform to the laminar solution (for $\Gamma = \infty$) and that the bulk has a much shallower slope---as seen in RB convection \cite{ahl09}---indicating turbulent mixing. For the laminar solution the convective part of $J_\omega$ is zero; any convection in the bulk of the system therefore decreases the $\partial_r \omega$ term, resulting in a shallower angular velocity gradient. From the insets we can clearly see that for increasing Taylor number the boundary layers become steeper and steeper, and indeed the angular velocity profile has then to become steeper as $u_r$ of eq.\ (\ref{jomega}) is zero at the wall, and we are only left with the term $\partial_r \omega$. We now split our data in two parts: $r' \in [0,1/2]$ (inner BL) and $r' \in [1/2,1]$ (outer BL) and normalize velocities with the appropriate $u_\tau$ (eqs.\ (\ref{equi}) and (\ref{equo})) and distances with the respective $\delta_\nu$ (eqs.\ (\ref{eqdeltai}) and (\ref{eqdeltao})), see figure \ref{fig:logprofiles}. We used \textit{global} torque measurements \citep{lew99,gil11} to find $u_\tau$ and $\delta_\nu$. In the viscous sublayer $(y^+ < 5)$ the velocity profiles are $u^+ = y^+$ which we can nicely resolve. It should be noted that global torque measurements provide an average torque $\tau$, while the local torque depends on height (following the large-scale Taylor vortex structure). So while eq.\ (\ref{eqtaui}) holds for the average $\tau_{\omega,i}$, it might have an axial dependence. To normalize the velocity profiles we used the globally measured torque and therefore we have the average of $\tau_{\omega,i}$, this causes the imperfect matching of $u^+=y^+$. The measurements were performed at middle height as sketched in fig.\ \ref{fig:setup}.

For $y^+>50$ the effects of viscosity diminish---indeed the contribution of $\nu \partial_r \omega$ to $J_\omega$ is very small in the bulk of the flow.  Furthermore, as suggested by Prandtl and von K\'arm\'an \cite{popeturb}, in this limit the velocity profile converges to:
\begin{align}
 u^+ &= \frac 1 \kappa \ln y^+ +B, \label{vonkarman}
\end{align}
with $\kappa$ the von K\'arm\'an constant and $B$ the logarithmic intercept. 

\begin{figure}[ht]
	\begin{center}
		\includegraphics{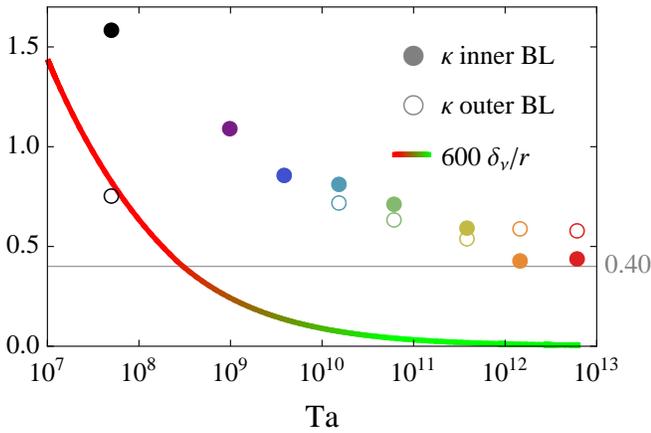}
		\caption{(color online) The parameter $\kappa$ obtained by fitting $u^+ = \frac 1\kappa \ln y^+ +B$ for $y^+ \in [50,600]$ near the inner and outer cylinder. The color scheme is identical to figures \ref{fig:fullprofiles} and \ref{fig:logprofiles}, the black symbols are DNS results \cite{ost12}. The solid colored line shows the ratio of the extent of the log-layer and the radius of the cylinder, indicating the influence of the curvature.}
		\label{fig:kappa}
	\end{center} 
\end{figure}

Indeed, outside the viscous wall region $y^+>50$ (inside the outer layer) figure \ref{fig:logprofiles} conclusively shows the existence of a log-layer in the ultimate TC regime ($\text{Ta} \gtrsim 10^9$), which is in sharp contrast to the laminar boundary layers found in the DNS simulations \cite{ost12} (indicated by thin lines) in the classical turbulent regime. Our profiles are fitted to eq.\ (\ref{vonkarman}) over the range $y^+ \in [50,600]$ (see the shaded area in figure \ref{fig:logprofiles}) and the values for $\kappa$ are extracted, see figure \ref{fig:kappa}. They are found to be different for the inner and outer BL and depend on $\text{Ta}$. For increasing Taylor number we see that the values for the inner BL approach a value of $\kappa \approx 0.40$, close to the known classical value of $\kappa = 0.40$ \cite{hultmark2012,meneveau2013}, recently systematically examined by Marusic \textit{et al.} \cite{marusic2013}. We included in figure \ref{fig:kappa} a line indicating the influence of the curvature in the log-regime ($600\delta_\nu/r$). It can be observed that for our lower Taylor numbers the importance of the curvature cannot be neglected, and that only for the very high Taylor numbers the curvature becomes negligible (green part of the solid line). For $\text{Ta}>10^{13}$ we expect the effect of curvature to become negligible and that the value for $\kappa$ approaches the classical value \cite{marusic2013}. We also would like to note that due to possible height dependence of $\tau_{\omega,i}$, and therefore the imperfect matching of $u^+ = y^+$, the values of $\kappa$ could slightly vary.

\begin{figure}[ht]
	\begin{center}
		\includegraphics{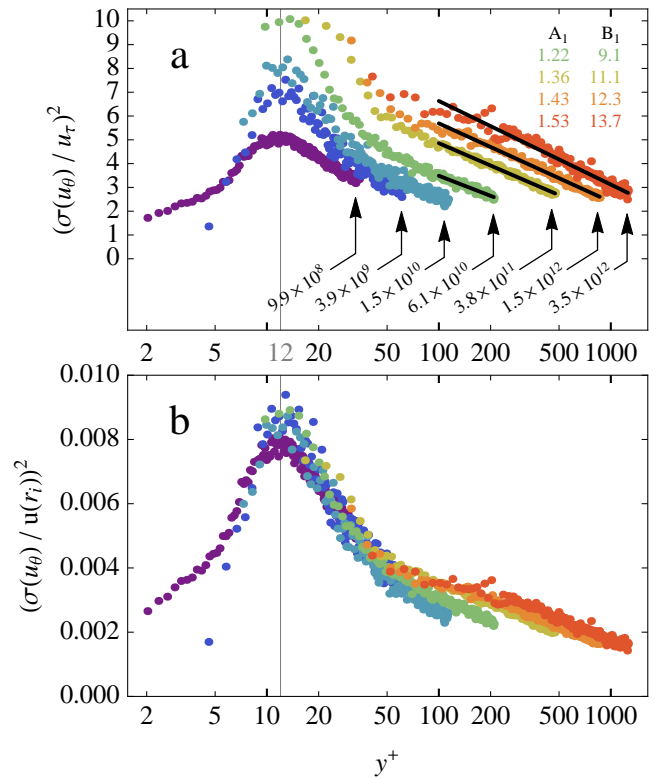}
		\caption{(color online)(a) The variance of the local azimuthal velocity is presented as a function of radial distance from the inner cylinder. The velocity is made dimensionless using the friction velocity $u_\tau$, and the distance is in wall-units $y^+ = \big( r - r_i \big)/\delta_\nu$. (b) Same as in figure a but the velocity has been rescaled using the driving velocity $u(r_i)$. Corresponding colors in figures a and b correspond to the same data-set, and are consistent with previous figures. The Taylor numbers are indicated with arrows in figure a. For $y^+>100$ the 4 highest $\text{Ta}$ cases are fitted with $(\sigma(u_\theta)/u_\tau)^2 = B_1 - A_1 \ln y^+$. The fitting parameters are listed inside figure a and colored accordingly.}
		\label{fig:sigma}
	\end{center} 
\end{figure}

In addition to PIV, also high resolution PTV measurements and analysis have been performed. For these measurement we zoomed into the area near the inner cylinder using a long-distance microscope to obtain a scaling factor of $\approx \unit{10}{\micro \meter \per px}$. The spatial resolution of PTV only depends on the number of images, and can thus be better than the pixel grid spacing projected in to physical space \cite{kah12a}. We extract the variance $\sigma^2(u_\theta)$ from the shape of the probability density function of the correlation function \cite{sch12}, see figure \ref{fig:sigma}. We normalize $\sigma(u_\theta)^2$ with the driving velocity $u(r_i)$ (see fig.\ \ref{fig:sigma}a) and with the friction velocity $u_\tau$ (see fig.\ \ref{fig:sigma}b). For both curves the maximum of $\sigma(u_\theta)^2$ is around $y^+ =12$, which is remarkably similar to the values obtained in pipe and channel flows (see \textit{e.g.} \cite{marusic2010,hultmark2012}). These peaks universally collapse inside the viscous wall region ($y^+ < 50$), where curvature effects do not play a role like in the log-regime. In addition it can be observed that indeed the collapse of the data is better when we normalize $\sigma(u_\theta)$ with the driving velocity rather than the shear velocity $u_\tau$. As shown in fig.\ \ref{fig:sigma}a, we fit the data for $y^+>100$ with $(\sigma(u_\theta)/u_\tau)^2 = B_1 - A_1 \ln y^+$ ---the log law for the velocity variance \cite{meneveau2013}. The corresponding fitting parameters are indicated in fig.\ \ref{fig:sigma}a. Remarkably, the slope $A_1$ (varying from $1.22$ to $1.53$) is comparable with values found ($A_1 \approx 1.25$) in high Reynolds number boundary layer flows \cite{meneveau2013}.


In conclusion, we performed direct measurements of the velocity boundary layer profiles in highly turbulent Taylor Couette flow up to $\text{Ta} = 6.2 \times 10^{12}$. In contrast to the laminar boundary layers in the classical turbulent TC regime, the present data in the ultimate regime provides direct experimental evidence of the emergence of a log-layer, as theoretically proposed in refs. \cite{kra62,gro11}. The fitted von K\'arm\'an constant $\kappa$ is found to approach the classical value of $0.40$ for large enough $\text{Ta}$. Furthermore, we find that the peak in the standard deviation of the azimuthal velocity universally collapses around $y^+ =12$, and that the height of the peak is found to collapse better when scaled with the driving velocity as compared to the friction velocity. Lastly, the variance profiles depict a log dependence for $y^+>100$.

\begin{acknowledgments}
We would like to thank Bas Benschop, Martin Bos, and Gert-Wim Bruggert for their technical support. We also acknowledge stimulating discussions with Bruno Eckhardt, Siegfried Grossmann, Ivan Marusic, Peter Monkewitz, and Roberto Verzicco. We thank Rodolfo Ostilla for making his DNS data available to us. This study was financially supported by the Simon Stevin Prize of the Technology Foundation STW of The Netherlands and the European Cooperation in Science and Technology (COST) Action program MP0806: ‘Particles in turbulence’. The high resolution PIV/PTV approaches were developed with financial support from the German Research Foundation (DFG) in the framework of the SFB-TRR 40 and the Individual Grants Programme KA 1808/8.
\end{acknowledgments}


\end{document}